\documentclass[a4paper,10pt,twoside]{cpc-hepnp}
\usepackage{CJK,upgreek,fancyhdr}
\usepackage{multicol}
\usepackage{graphicx}
\usepackage{booktabs}
\usepackage{amssymb,bm,mathrsfs,bbm,amscd}
\usepackage[tbtags]{amsmath}
\usepackage{lastpage}
\usepackage{epstopdf}

\makeatletter

\newcommand{\Rmnum}[1]{\expandafter\@slowromancap\romannumeral #1@}
\makeatother

\begin{document}
%\begin{CJK*}{GB}{gbsn}
\begin{CJK*}{GBK}{song}

\fancyhead[c]{\small Submitted to Chinese Physics C}
\fancyfoot[C]{\small 010201-\thepage}

\footnotetext[0]{Received 31 June 2015}

\title{On-Line Cluster Reconstruction Of GEM Detector Based On FPGA Technology\thanks{Supported by National Natural Science
Foundation of China (Grants  No. 11405077, No.11575073 and No.11575268) }}
\author{  Hui-Yin Wu$^{1}$
\quad He-Run Yang$^{2}$
\quad Wei  Zhang$^{1}$
\quad Jian-Jin Zhou$^{1}$
\quad  Sheng-Ying Zhao$^{1}$ \\
\quad  Chen-Gui Lu$^{2}$
\quad Jun-Wei Zhang$^{2}$
\quad  Bi-Tao Hu$^{1}$
\quad Yi Zhang$^{1;1}$\email{yizhang@lzu.edu.cn}
\quad Wen-Quan Cao$^{3;2}$\email{caowenquan@nwnu.edu.cn}
}
\maketitle

\address{%
$^1$ School of Nuclear Science and Technology, Lanzhou University, Lanzhou 730000, China\\
$^2$ Institute of Modern Physics, Chinese Academy of Sciences, Lanzhou, 730000, China\\
$^3$ College of Computer Science \& Engineering Northwest Normal University, Lanzhou, 730000, China
}

\begin{abstract}
In this work, a serial on-line cluster reconstruction technique based on FPGA technology was developed to compress experiment data and reduce the dead time of data transmission and storage. At the same time, X-ray imaging experiment based on a two-dimensional positive sensitive triple GEM detector with an effective readout area of 10 cm $\times$ 10 cm was done to demonstrate this technique with FPGA development board. The result showed that the reconstruction technology was practicality and efficient. It provides a new idea for data compression of large spectrometers.
\end{abstract}



\footnotetext[0]{\hspace*{-3mm}\raisebox{0.3ex}{$\scriptstyle\copyright$}2013
Chinese Physical Society and the Institute of High Energy Physics
of the Chinese Academy of Sciences and the Institute
of Modern Physics of the Chinese Academy of Sciences and IOP Publishing Ltd}%

\begin{multicols}{2}

\section{Introduction}
GEM detectors are used to detect charged particles due to their excellent performances (high gain, good spatial resolution, and high count rate)~\cite{lab1,lab2}, especially in high energy nuclear physics experiments. As a track detector, it is required a high spatial resolution. Standard GEM detector readout board has 512 signal channels for a 10cm$\times$10cm effective area. For a larger area of the GEM detector, there will be more signal channels~\cite{lab3,lab4,lab5}. Large-scale GEM detector requires a support of front-end electronics in a huge scale, especially in the spatial resolution demanding experiment~\cite{lab6}. In nowadays, most of the integrated pre-amplified data acquisition system is based on the FPGA (Field-Programmable Gate Array) technology for data processing and transmission~\cite{lab7,lab8}. When the detector detects the radiation, signals on every electrode will be amplified with integrated pre-amplifier, and transferred to the storage medium using FPGA based on a data bus such as CAMAC or VME, regardless of whether the data is useful~\cite{lab7}. However, the transmission speed of these protocols is too slow compared with the intrinsic dead time of a two-dimension spatial sensitive GEM detector, which can tolerate very high count rates. The speed of data transmission and storage limits the maximum counting rate of the whole system. In most cases, only are the real signals on a small part of all electrodes, while the others can be discarded reasonably. We propose a method of compressing the data in the FPGA chip based on the programmability of FPGA. In this method, only the useful signals are transmitted while other signals are discarded to reduce the dead time due to data transmission and storage.

\section{On-line cluster reconstruction}
FPGA as a programmable logic gate device, has a very good flexibility, not only can complete the data transfer function but also has some complex data processing functions. Considering that the data acquisition system in most nuclear experiment is based on FPGA technology, so we proposed to use FPGA to implement on-line cluster reconstruction and data compression.
\\
\indent
In high-energy nuclear physics experiment with large beam flux, the detector will simultaneously detect multiple incident particles, and these particles will deposit energy in different locations of the detector. We can not simply reconstruct the signals of readout electrons as one cluster with central method, but rather reconstruct clusters from different locations independently. How to distinguish the signals generated in the same time but different locations quickly and accurately is the key to achieving cluster reconstruction and data compression.

\subsection{Reconstruction principle}

The 12-bits data of detector readout electron from ADC flow from left to right into the reconstruction module as shown in Fig.~\ref{fig1}. At the beginning of the module processing, 12-bits data will be compressed into 1-bit data by compared with noise threshold, for saving up chip resources. Normally, if the data is larger than the noise threshold, this electron will be marked as 1. Otherwise it will be marked as 0. When checking the electron labelled 1 in Fig.~\ref{fig1}, the electron  marked black have been checked already and the electron marked orange have not. If the electron is fired while all of its checked neighbors are unfired, the electron is identified as a new signal cluster. To the contrary, if the electron has least one neighbor is checked as fired, the electron is grouped into the same signal cluster as this fired neighbors. In some cases, two electrons belonging to the same cluster maybe misjudged to be two different clusters, as Cluster \Rmnum{4} and Cluster \Rmnum{5} showde in Fig.~\ref{fig1}. In this case, electron 12 is the key to fixing the bug. When the module judges the four neighbor electrons around electron 12, it will find the cluster of left electron and the cluster of lower right is not the same cluster. Obviously, they are same one. The module will mark this two clusters, and all the marked clusters will be re-merged after the electrons are assigned to the respective cluster. After all, $\sum$$X\cdot Q_x$, $\sum$$Y\cdot Q_y$, $\sum$$Q_x$, $\sum$$Q_y$, $\sum$$X$, $\sum$$Y$, $X_{min}$, $X_{max}$, $Y_{min}$, $Y_{max}$ of each cluster will be stored in the FIFO and wait to be transmitted.
\begin{center}
\includegraphics[width=7cm]{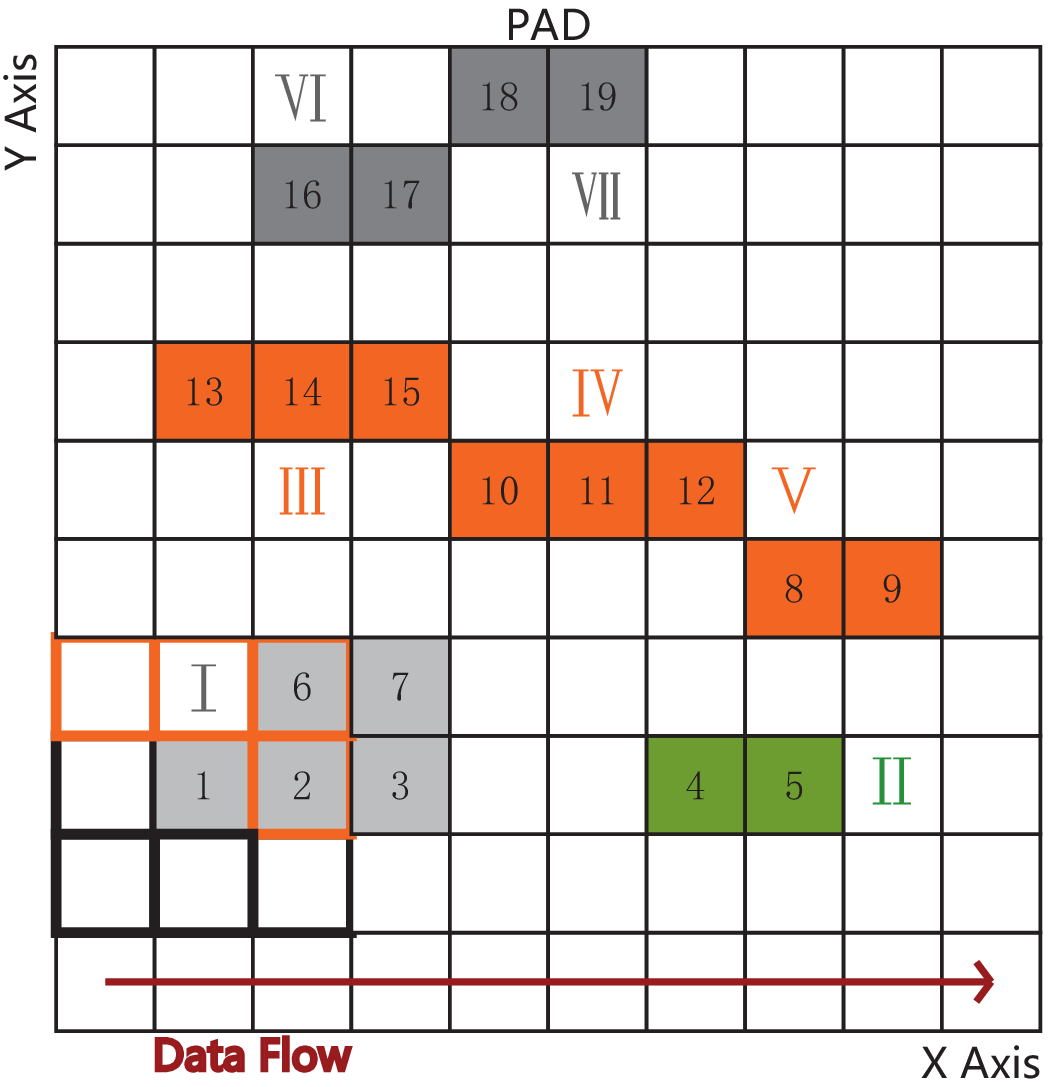}
\figcaption{\label{fig1}(Color Online) Schematic view of signal cluster algorithm. The boxes filled with color mean that the electrodes are fired, while the empty boxes mean the electrodes are not. Arabic numeral on the electron represents the order in which the data enters the module.  }
\end{center}
\subsection{Reconstruction test}
A complex case shown in Fig.~\ref{fig2} is tested by the reconstruction module with an FPGA development board(Altera DK-DEV-2AGX125N). The result showed that all the clusters including square shape, strip shape and X shape could be reconstructed accurately. All the information of each cluster is also given by the reconstruction module correctly.

  \begin{center}
\includegraphics[width=8cm]{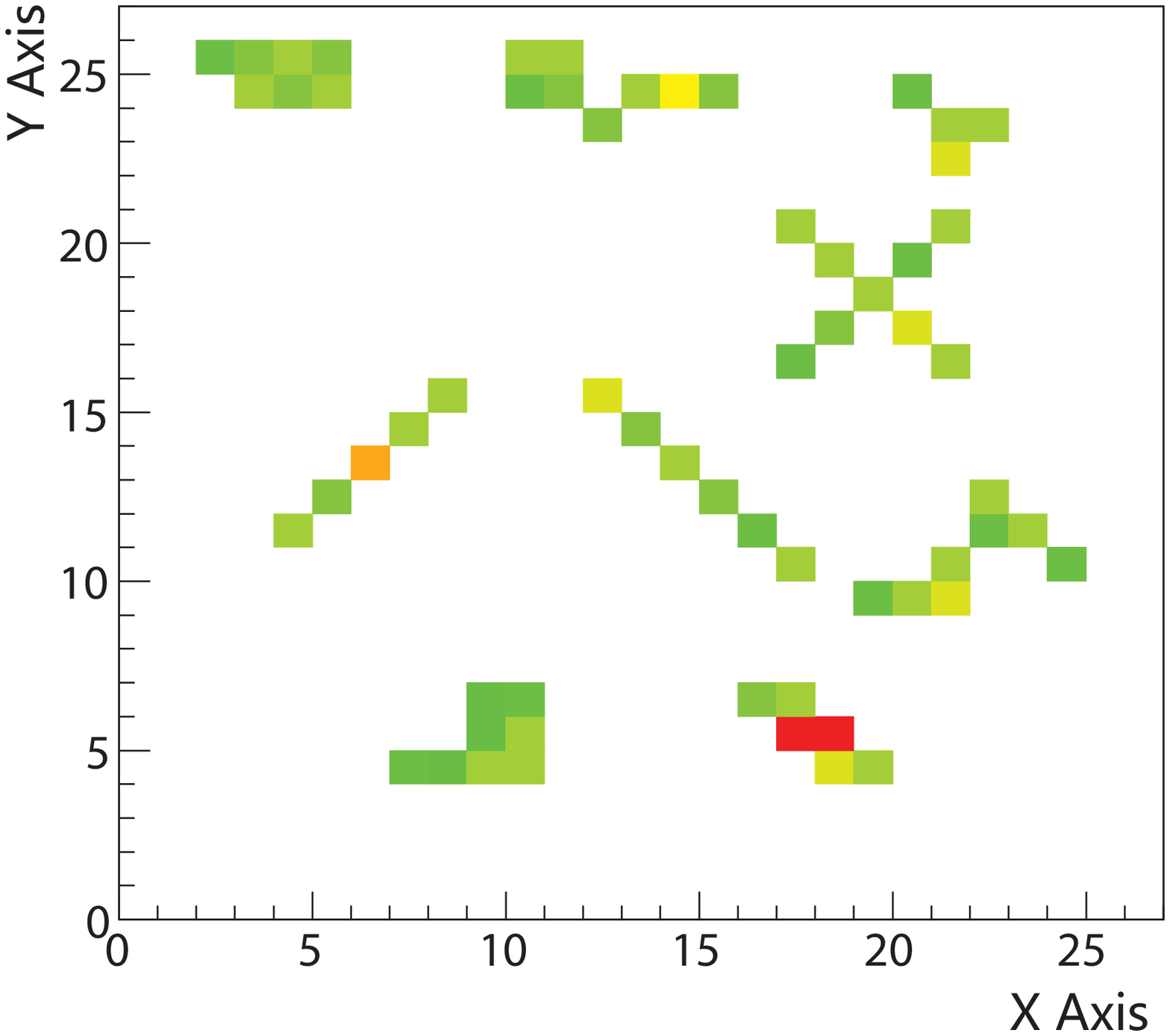}
\figcaption{\label{fig2} Reconstruction test. A configuration of complex case with various clusters is tested by the reconstruction module. The color represents the amount of charge.}
\end{center}
\subsection{Time consumption}
Time consumption is a very important indicator because reconstruction method is based on a serial processing. The FPGA development board tested the time consumption under different duty ratio shown in Fig.~\ref{fig3}. In the non-extreme case (The ratio is about 20\% $\sim$ 30\%), the processing cycle is five times of the data input period. This means that a large-capacity FIFO is required for temporary storage of data. The use of frequency FIFO can solve the problem of longer cycle consumption.
 \begin{center}
\includegraphics[width=8cm]{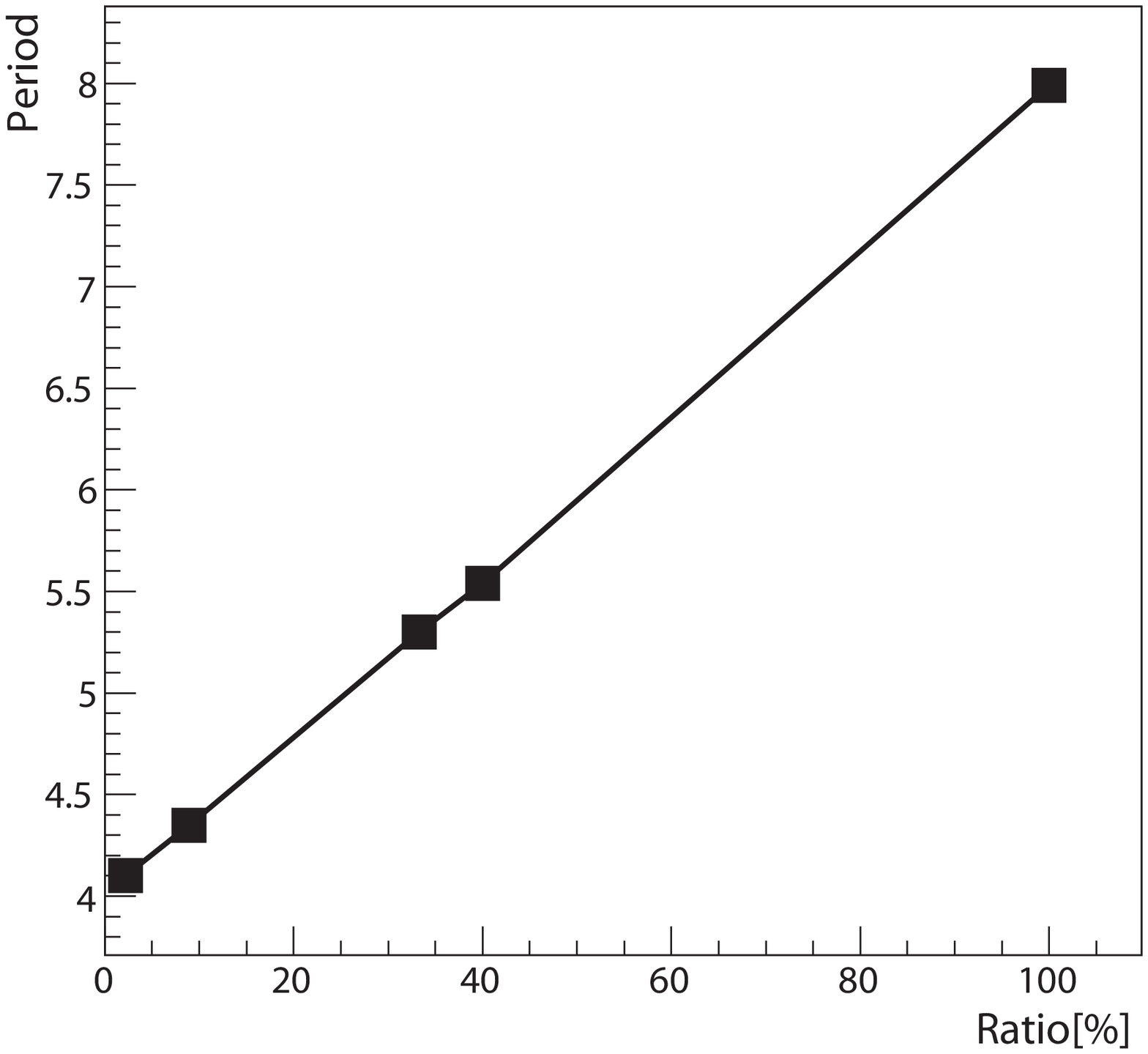}
\figcaption{\label{fig3}  Time consumption of reconstruction module. The X-axis represents the ratio of the number of electron above the noise threshold to the total electron number, while Y-axis represents the ratio of the time taken by the module to reconstruct clusters to the total time of the signal enters the module. }
\end{center}
\section{Experimental test}
The experimental data of X-ray measured by a two-dimensional sensitive GEM detector is used to test the reconstruction module.
\subsection{Detector assembly}
The two-dimensional  sensitive detector, which consists of a cathode plane, three GEM foils, and a read-out anode works in a proportional mode as shown in Fig.~\ref{fig4}. The detector has a square active area of 100 mm $\times$ 100 mm, a 3mm drift gap, a 4mm induction, and two 2 mm transfer gaps. The detector is operated based on a continuously flushed Ar/CO$_2$ gas mixture (80/20 percentage in volume). A voltage divider was employed to supply the bias voltage to the detector, which avoids electric fields becoming too high in the case of  discharge. For further protection, an additional 10 M$\Omega$ protection resistor was connected with the electrode. When the electron is avalanched by the third GEM foil, the voltage of this foil will be reduced. The lower film signal of the third layer GEM film led out by a blocking capacitor and amplified by a pre-amplifier can be used as a trigger for the DAQ system.
\begin{center}
\includegraphics[width=8cm]{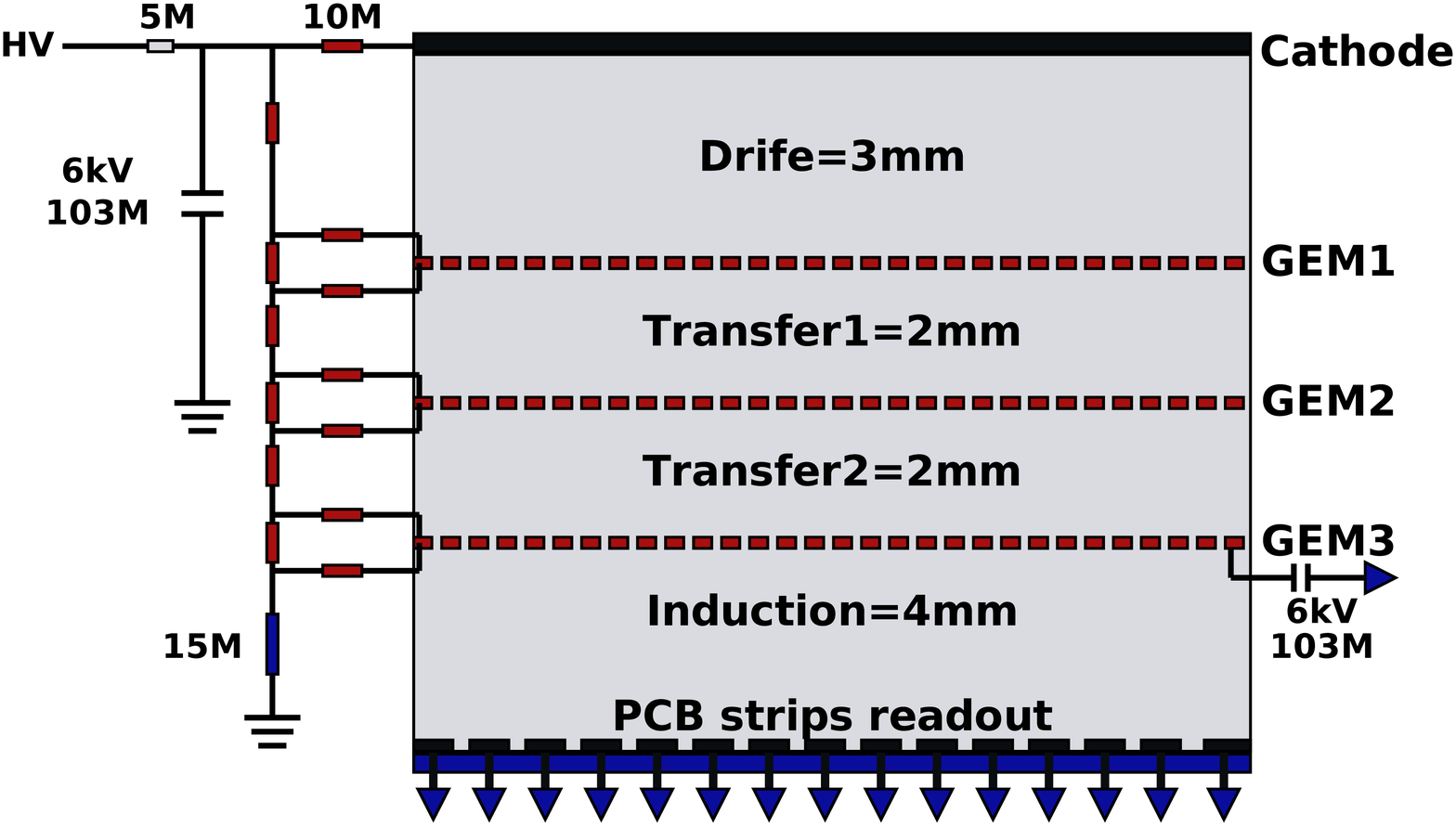}
\figcaption{\label{fig4}   Experimental setup of triple GEM detector. The lower film signal of the third layer GEM film led out by a blocking capacitor and amplified by a pre-amplifier to be used as a trigger for the DAQ system }
\end{center}
\par
\indent
The anode plane is made of two-dimensional parallel strips with standard PCB (Printed Circuit Board) technology. The pitch is 600 $\upmu$m for all read-out strips (167 strips for 100 mm), the strip width of X-axis is 130 $\upmu$m  and Y-axis are composed of 280 $\upmu$m $\times$ 280 $\upmu$m squares connected through in the third layer by the strip(167 strips for 100 mm). The specific size of the read-out PCB is showed in Fig.~\ref{fig5}. All avalanche electrons can be absorbed by the strips in both dimensional equally because the strip areas of both dimensional are the same.

\begin{center}
\includegraphics[width=7cm]{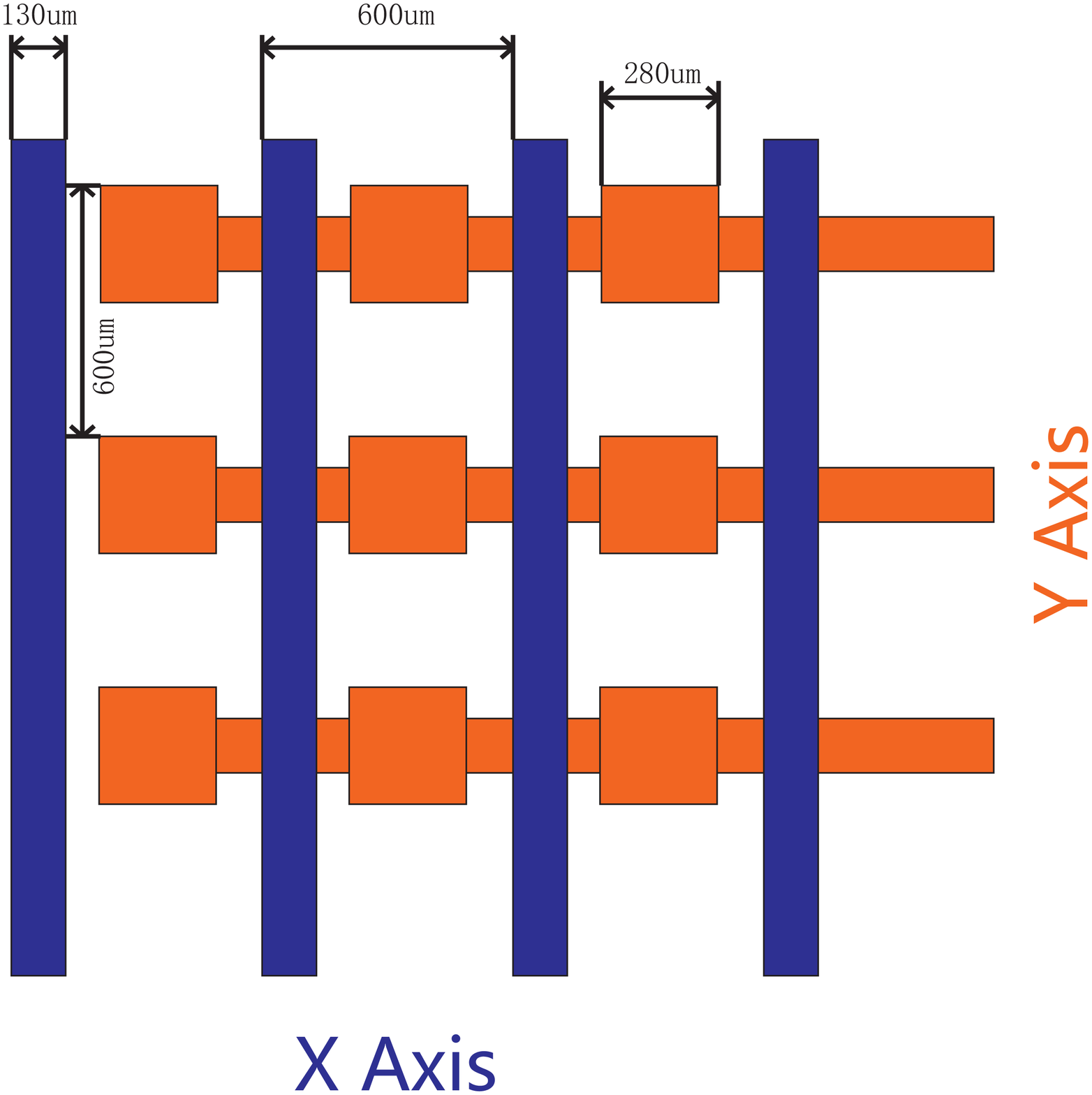}
\figcaption{\label{fig5}    Readout board PCB schematic. The readout area of the X-axis is equal to the read-out area of the Y-axis so that the charge can be evenly distributed in both dimensions, ensuring that spatial resolution of both dimension is equivalent.   }
\end{center}
\subsection{Energy resolution}
The energy resolution of the triple GEM detector was measured with $^{55}$Fe 5.9 keV X-ray source, as shown in Fig.~\ref{fig6}. We added signals of the valid strips (3 $\sim$ 4 strips) as a whole signal. The energy resolution is about 18.7\%.
\begin{center}
\includegraphics[width=8cm]{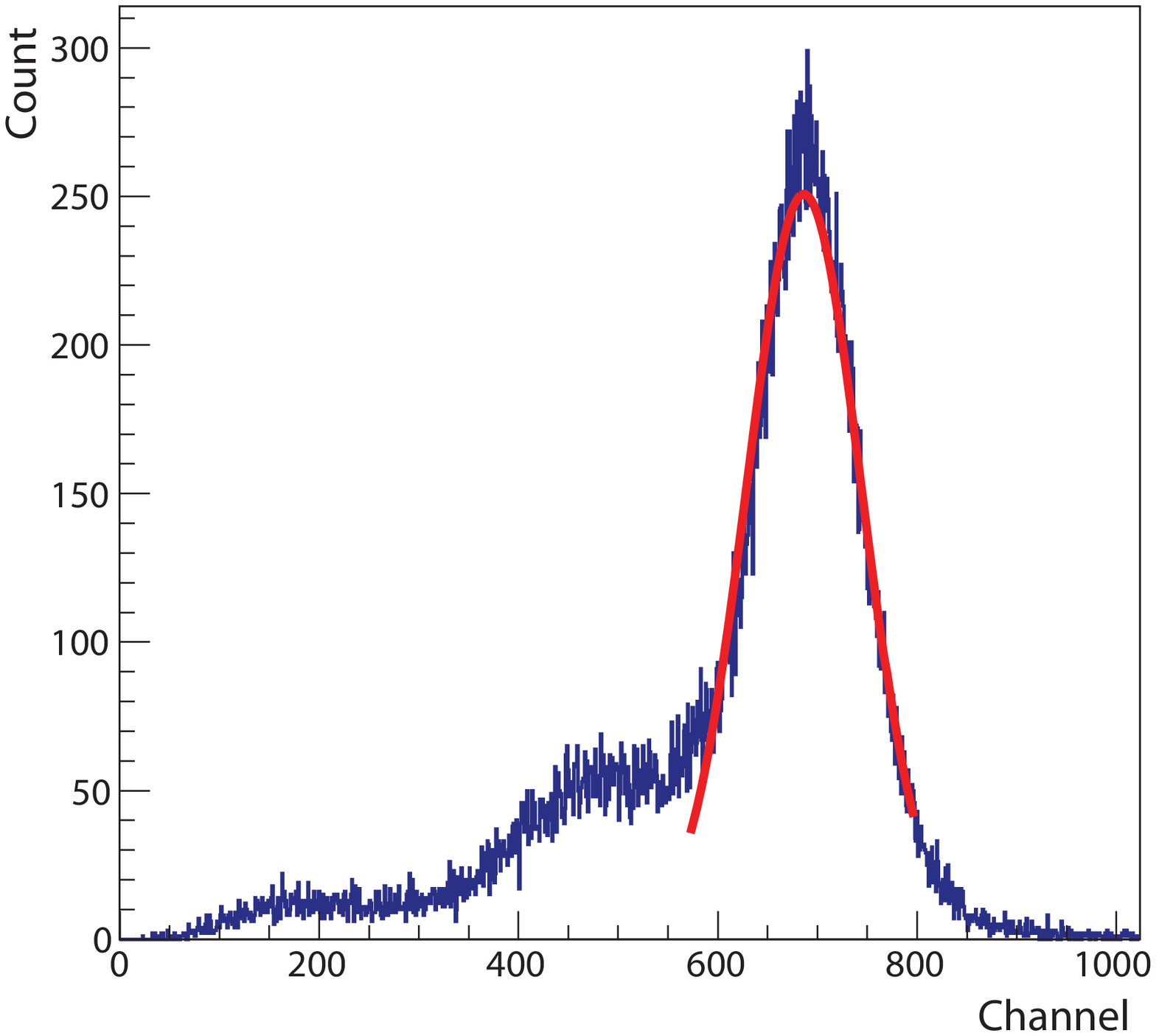}
\figcaption{\label{fig6} (Color online) $^{55}$Fe 5.9 keV X-ray source spectrum measured by triple GEM detector }
\end{center}
\subsection{Spatial resolution}
The focus of this work is on-line cluster reconstruction, so a simple measurement of spatial resolution is tested by a $^{55}$Fe X-ray source with a 1 mm width slit. The convolution fit method~\cite{lab9} is used to measure the spatial reconstruction shown in FIG.~\ref{fig7}. The spatial resolution is about 127 $\upmu$m.
\begin{center}
\includegraphics[width=8cm]{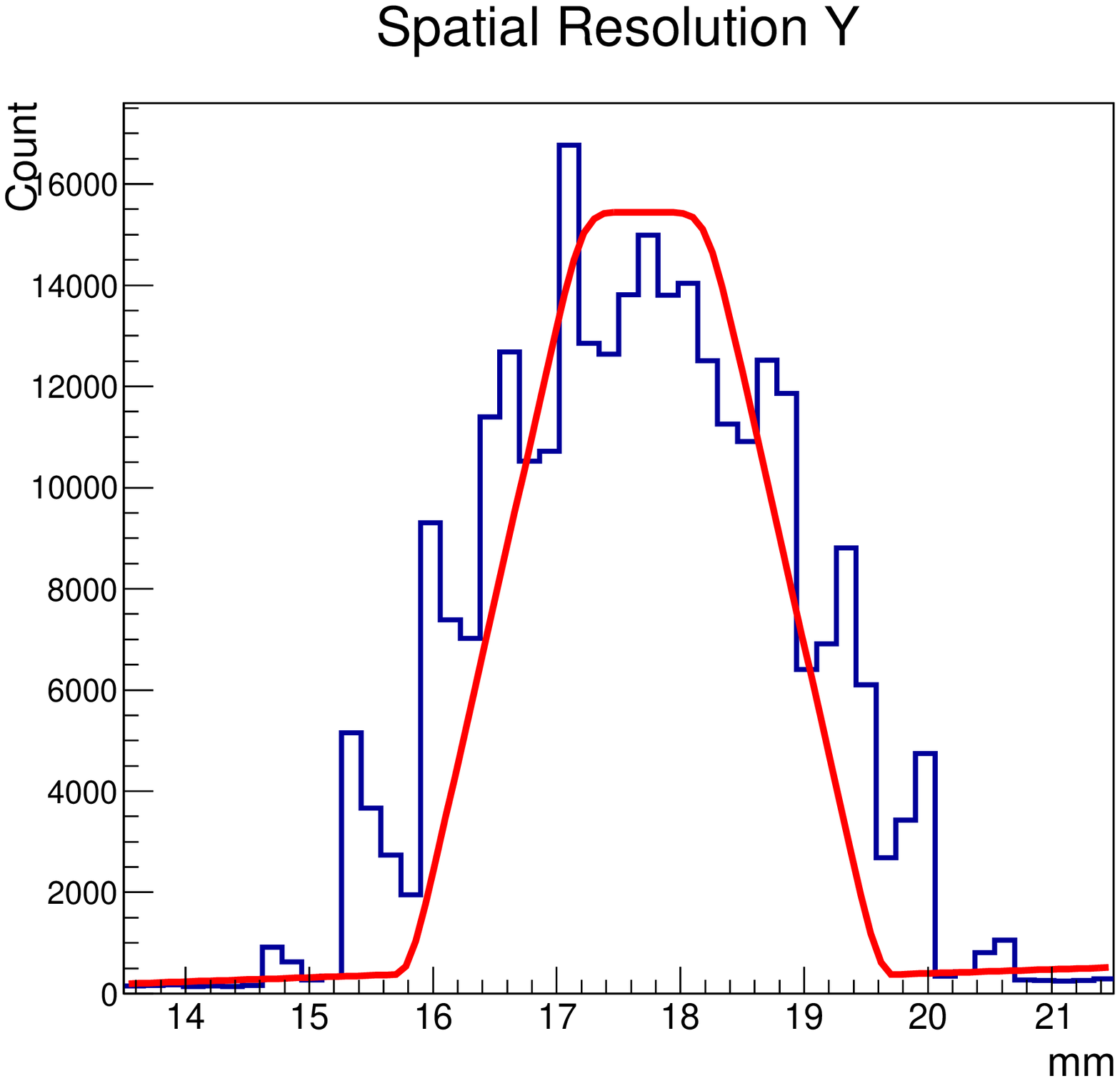}
\figcaption{\label{fig7} (Color online) Spatial spectrum of triple GEM detector. }
\end{center}
\subsection{Imaging experiment and data process}
The badge of Lanzhou University, made by 3D printing technology, was imaged with a high activity $^{55}$Fe X-ray source. The radiation source was 45 cm away from the imaging object, and the imaging object was placed on the entrance window of GEM detector. An X-ray will fire 3-4 strips in one-dimensional as shown in Fig.~\ref{fig8}, and the position is determined by Gaussian fitting. The final imaging result is showed in Fig.~\ref{fig9}.
 \begin{center}
\includegraphics[width=8cm]{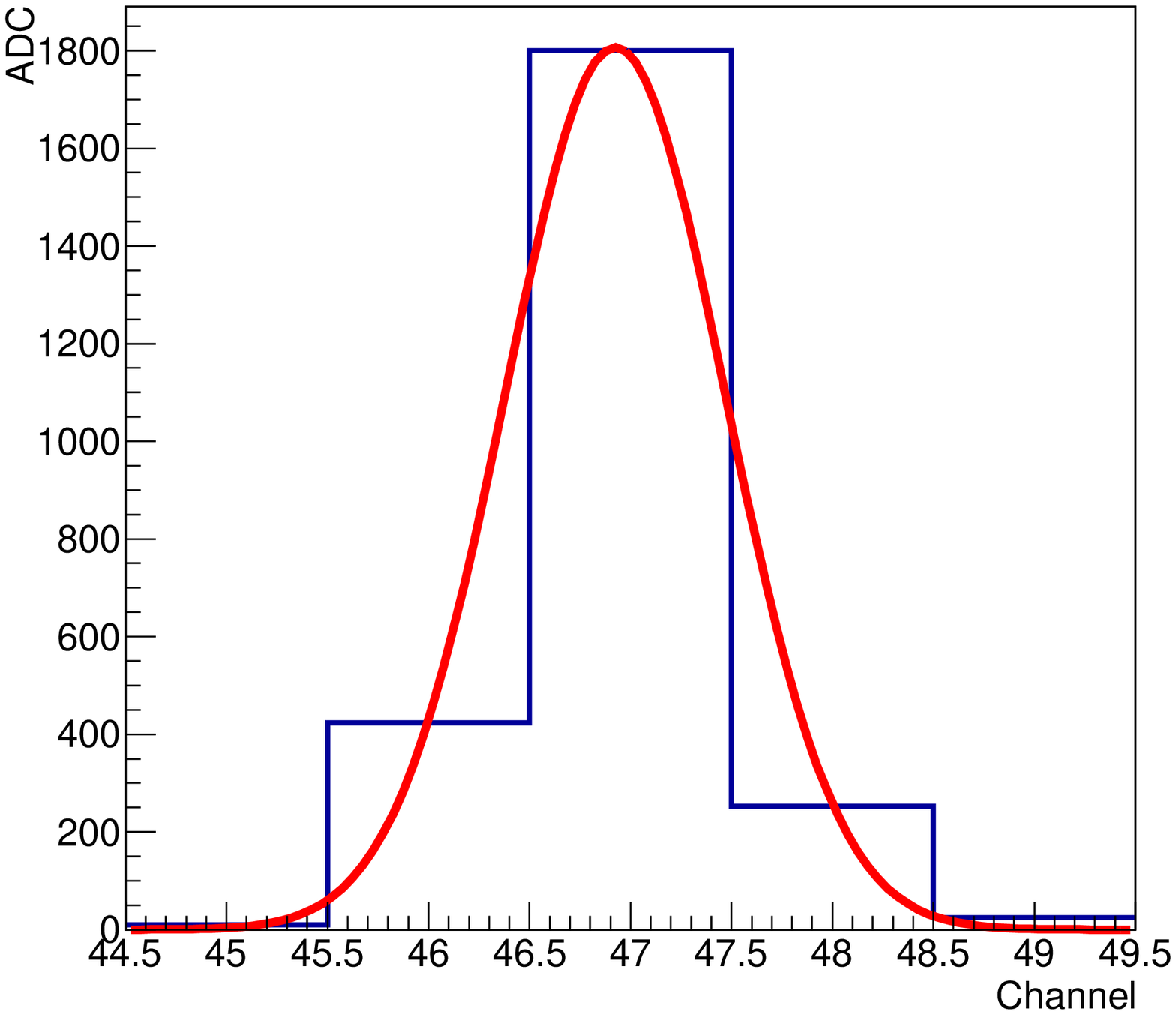}
\figcaption{\label{fig8}    The centre value of Gaussian Fitting is the incident position of X-ray }
\end{center}
 \begin{center}
\includegraphics[width=8cm]{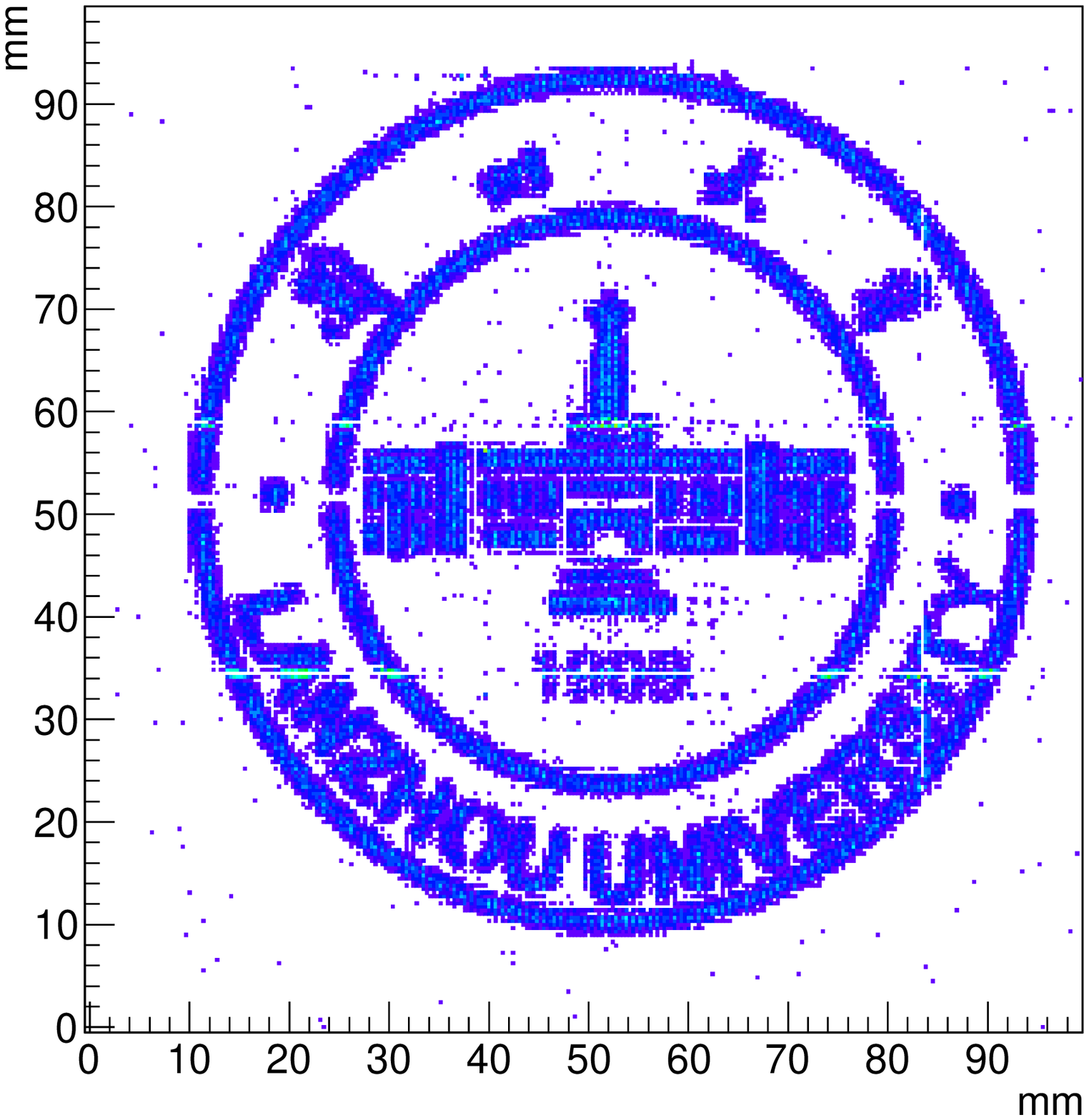}
\figcaption{\label{fig9}   Imaging result of the badge of Lanzhou University. The size of the badge is 80mm $\times$ 80mm, which smaller than the effective area of the detector.  }
\end{center}

\subsection{Reconstruction with experiment data}
Due to the limitation of readout electronics, the experiment used the strip readout construction, the strip readout signal will be treated as 1 $\times$ 167 electrons readout construction in one direction during reconstruction experimental data.  Each signal of X-ray is reconstructed by the reconstruction module with FPGA development board,  $\sum$$X\cdot Q_x$, $\sum$$Y\cdot Q_y$, $\sum$$Q_x$, $\sum$$Q_y$, $\sum$$X$, $\sum$$Y$ of each cluster is read from FPGA development board. The incidence position of each X-ray is determined by the centre of gravity method, and the imaging result is showed in Fig.~\ref{fig10}. Lanzhou University badge can still be clearly visible, which proves the on-line cluster reconstruction technology can achieve fast cluster reconstruction and data compression.

 \begin{center}
\includegraphics[width=8cm]{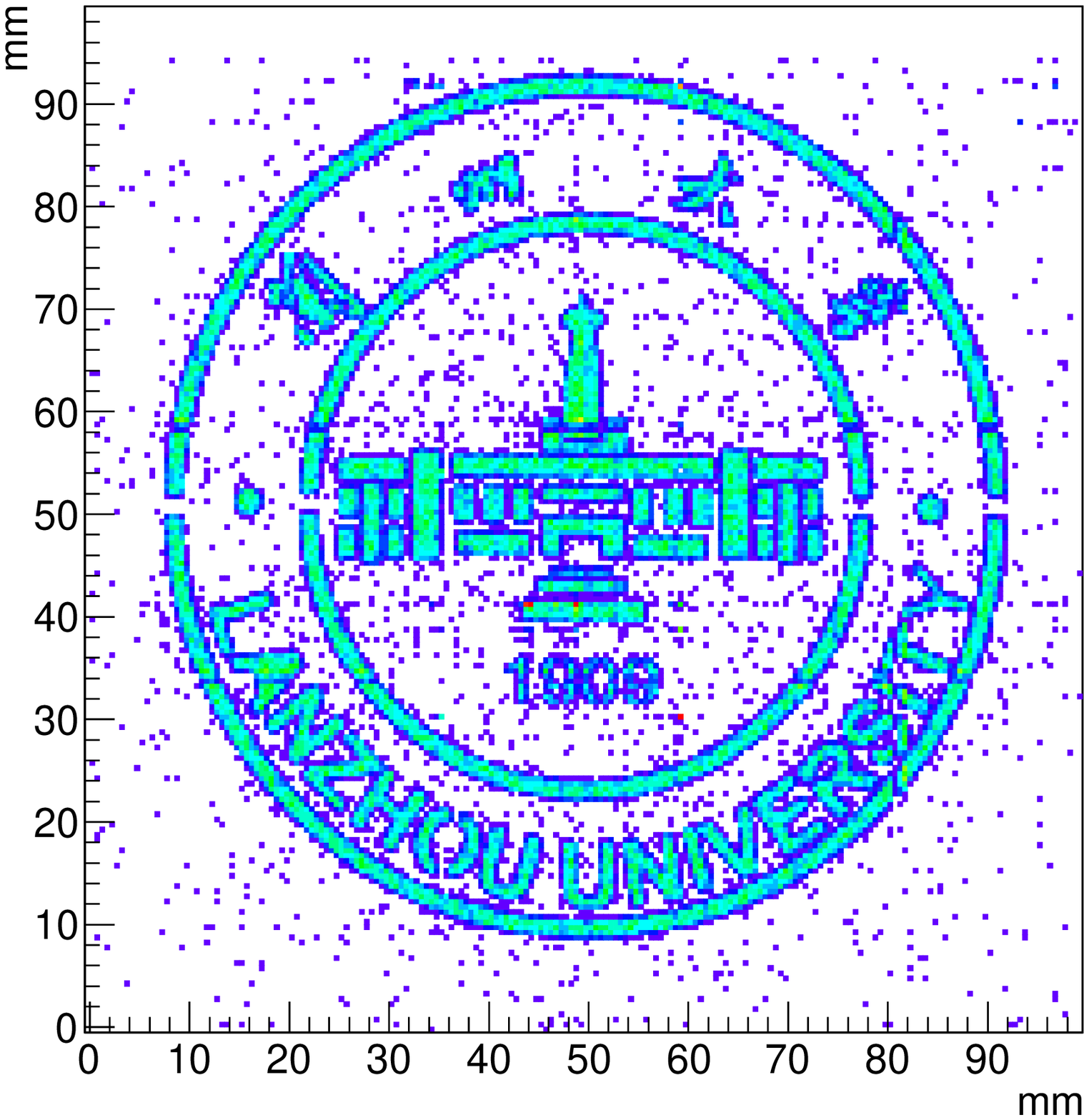}
\figcaption{\label{fig10}   Imaging result of the badge of Lanzhou University badge processed by FPGA development board with cluster reconstruction module. }
\end{center}

\section{Conclusions}
The FPGA development board is used to develop an on-line cluster reconstruction technology based on FPGA technology, and the X-ray imaging experiment of Lanzhou University badge with a two-dimensional position sensitive GEM detector shows that this technology is feasible and high-efficiency and data compression is effective. Reconstruction algorithm should be carefully tested and optimised, because the raw experiment data will be deleted permanently. Further development, online cluster reconstruction can into online track reconstruction with time information, and it can correct the incident points of neutrons by reconstructing each cycle of the track using the time information of recoil protons~\cite{lab10}, improving the precision of neutron imaging, and achieving neutron imaging quickly~\cite{lab11,lab12}.

\par
\indent
However, our algorithm has been able to work very well, it provides a new idea for data compression of large spectrometers and reduces the dead time due to data transmission.
\end{multicols}

\vspace{-1mm}
\centerline{\rule{80mm}{0.1pt}}
\vspace{2mm}

\begin{multicols}{2}

\end{multicols}

\clearpage
\end{CJK*}

\begin{thebibliography}{90}

\vspace{3mm}

\bibitem{lab1} F.Sauli, Nucl. Instrum. Methods A, 386(2): 531-534(1997)

\bibitem{lab2} A. Bressan et al, Nucl. Instrum. Methods A, 425: 254-261(1999)

\bibitem{lab3} F.Sauli, Nucl. Instrum. Methods A, 805: 2-24(2016)
\bibitem{lab4} Abbaneo D, Abbas M, Abbrescia M, et al. Nucl. Instrum. Methods A, 805: 2-24(2016)
\bibitem{lab5} Gnanvo K, Liyanage N, Nelyubin V, et al. Nucl. Instrum. Methods A, 782: 77-86(2015)
\bibitem{lab6} Bamberger A, Desch K, Renz U, et al.Nucl. Nucl. Instrum. Methods A, 573(3): 361-370(2007)
\bibitem{lab7} Basile E, Bellini V, Capogni M, et al. Nucl. Instrum. Methods A, 718: 429-431(2013)
\bibitem{lab8} French M J, Jones L L, Morrissey Q, et al. Nucl. Instrum. Methods A, 466(2): 359-365(2001)
\bibitem{lab9} Xin-Yu L, Rui-Rui F, Yuan-Bo C, et al. Chin. Phys. C, 36(3): 228(2012)
\bibitem{lab10} Xiao-Dong W, He-Run Y, Zhong-Guo R, et al. Chin. Phys. C, 39(2): 026001 (2015)
\bibitem{lab11} Huang M, Li Y, Deng Z, et al.  IEEE NSS/MIC,1-3(2013)
\bibitem{lab12} Yi Z, Xiao-Dong Z, Wen-Xin W, et al. Chin. Phys. C, 33(1): 42(2013)

\end{thebibliography}
\end{document}